# Justice Blocks and Predictability of U.S. Supreme Court Votes

Roger Guimerà[1,2]*, Marta Sales-Pardo[2]*

1 Institució Catalana de Recerca i Estudis Avançats (ICREA), Barcelona, Catalonia, Spain, 2 Departament d'Enginyeria Química, Universitat Rovira i Virgili, Tarragona, Catalonia, Spain

**Abstract**

Successful attempts to predict judges' votes shed light into how legal decisions are made and, ultimately, into the behavior and evolution of the judiciary. Here, we investigate to what extent it is possible to make predictions of a justice's vote based on the other justices' votes in the same case. For our predictions, we use models and methods that have been developed to uncover hidden associations between actors in complex social networks. We show that these methods are more accurate at predicting justice's votes than forecasts made by legal experts and by algorithms that take into consideration the content of the cases. We argue that, within our framework, high predictability is a quantitative proxy for stable justice (and case) blocks, which probably reflect stable a priori attitudes toward the law. We find that U.S. Supreme Court justice votes are more predictable than one would expect from an ideal court composed of perfectly independent justices. Deviations from ideal behavior are most apparent in divided 5–4 decisions, where justice blocks seem to be most stable. Moreover, we find evidence that justice predictability decreased during the 50-year period spanning from the Warren Court to the Rehnquist Court, and that aggregate court predictability has been significantly lower during Democratic presidencies. More broadly, our results show that it is possible to use methods developed for the analysis of complex social networks to quantitatively investigate historical questions related to political decision-making.

**Funding:** This work was supported by Spanish Ministerio de Ciencia e Innovación (MICINN) Grant FIS2010-18639, James S. McDonnell Foundation Research Award, European Union Grant PIRG-GA-2010-277166 (to R.G.), and European Union Grant PIRG-GA-2010-268342 (to M.S.-P.). The funders had no role in study design, data collection and analysis, decision to publish, or preparation of the manuscript.

**Competing Interests:** The authors have declared that no competing interests exist.

* E-mail: roger.guimera@urv.cat (RG); marta.sales@urv.cat (MS-P)

## Introduction

Could we replace a justice of the U.S. Supreme Court by an algorithm that does not know anything about the law or the case at hand, but has access to the remaining justices' votes and to the voting record of the court? Surely, the algorithm would not be able to hear the case or write an opinion, but would it be able to mimic the vote of the missing judge? Here, we investigate these questions and discuss the implications for quantifying deviations from "ideal" judicial behavior and for getting quantitative insights into the dynamics and historical evolution of the judiciary.

Questions of justice predictability have deep implications from at least two perspectives: the perspective of decision theory and the perspective of legal studies and political science. From a decision–theory perspective, Supreme Court decisions are singular because, in some ways, they are expected to be quasi–ideal. Indeed, justices are expected to make rational decisions with almost perfect information, that is, as perfect as one may possibly expect in any complex real–world decision–making situation. Additionally, the decision of one justice may be expected not to affect the decisions of the others, that is, the decisions may be expected to be non–strategic in a game–theoretic sense [1]. Despite all of this, unanimous decisions are not the norm; and even when decisions of the Court are unanimous, legal experts have difficulties in predicting, not only that the vote will be unanimous, but even the "sign" of the Court's decision–in a study of the 2002 term of the Supreme Court, experts incorrectly predicted the global outcome of as many as 34.7% of the Court's unanimous decisions [2]. Even more puzzling, although consistent with empirical evidence from real decision–making processes, are recent observations that extraneous factors (such as the position of a case within a session) affect judges' decisions because of psychological biases [3,4].

From the perspective of legal studies and political science, on the other hand, scholars have argued that, by trying to predict the behavior of judges, one can get insights into how legal decisions are truly made [2,5,6]. This argument usually appears in the framework of the "debate" between legalism and attitudinalism (or, more broadly, legal realism), which argue, respectively, for purely legal versus personal attitude explanations to justices' decisions [7]. Within this debate, successful predictions based on variables describing the case under consideration (such as the issue area of the case or the ideological direction of the lower court ruling) lend evidence to one theory or the other.

Here we take an approach that is complementary to these vote–forecasting efforts. Specifically, we investigate whether and to what extent it is possible to make predictions of a justice's vote based on the other justices' votes in the same case (and the track record of the court), independently of case content. To make our predictions, we use methods that have been developed to analyze complex social networks [8,9] and complex affiliation networks [10,11], and that are able to uncover hidden associations between social actors (in this case, justices) [12]. We show that our approach is more accurate at predicting justice's votes than forecasts made by legal experts and by algorithms that take into consideration the content of the cases [2,5].

Moreover, we argue that a justice's predictability relative to her predictability in an equivalent "ideal court" provides a quantitative proxy for stable voting blocks (groups of justices that vote consistently the same way in groups of cases) [13–16], which ultimately reflect stable a priori attitudes towards the law. We find that Supreme Court justices are significantly more predictable than one would expect from "ideally independent" justices in "ideal courts" [17]. Deviations from ideal behavior are most prominent in divided 5–4 decisions, where justice blocks seem to be most stable. We find evidence that different justices have significantly different relative predictabilities (which suggests that some justices are more stably associated to certain justice–case blocks than others) and that justice relative predictability has decreased during the 50–year period spanning from the Warren Court to the Rehnquist Court. We also find that courts, as a whole, have been significantly less predictable during Democratic presidencies.

### Court idealizations

In an "omniscient Court" [17], justices are perfectly rational and free of preferences or attitudes, and have access to complete information about the case at hand. Given these legalist idealizations, all justices must reach the same conclusion and all cases must result in unanimous decisions [17]. Of course, the omniscient Court model is trivially refuted by the empirical observation that not all cases result in unanimous decisions. Therefore, one must relax some of the assumptions to account for justice variability.

Still within the legalist idealization, one may assume that the information available to justices is not perfect and/or that there is some other source of uncertainty. In such an "ideal Court", each justice evaluates, still free of ideology and independently of other justices, the merits of a case taking into account, to the best of her abilities and knowledge, law and precedent. Cases, on the other hand, all raise rigorously new issues, so previous decisions cannot "easily" determine by themselves the outcome of the present case. Without this assumption that each new case is intrinsically hard, one might think, for example, of a situation where all cases are identical and each justice votes the same every time, even in an "ideal Court." The assumption of intrinsic difficulty of the cases seems relatively weak given that: (i) the Supreme Court is mostly an appellate court, with ultimate and discretionary appellate jurisdiction over state and federal courts; (ii) even legal experts have difficulties predicting court decisions [2].

Under these assumptions, the ruling on a case can be modeled as a binomial process where each justice has the same probability $q$ of agreeing with the petitioner, so that "easy cases" (those with $q \approx 0$ or $q \approx 1$) result in unanimous decisions, whereas "hard cases" ($q \approx 0.5$) generally result in divided votes. The defining characteristic of an ideal court is that justices' votes are uncorrelated, that is, the fact that two justices agree (or disagree) on one case carries no information about their potential agreement on another case.

Because of the lack of correlations, the best possible algorithm to predict the vote of a justice in an ideal court, given the vote of the other eight, is the majority rule (one could in general do better than the majority by estimating, from previous cases, the distribution of $q$ values; in practice, this requires more than 150 cases and is impractical) [18]: if the majority of the eight justices agreed with the petitioner, predict agreement; if the majority disagreed, predict disagreement; in case of a tie, toss a coin.

If the court is less than ideal and some of its justices cast votes with a consistent bias, the decisions of individual justices become more predictable because, given the vote of eight justices on a case, one can use the track record of the court to classify the case into a certain "block"; then the track record of the ninth justice enables one to assess what is her most likely decision for cases in that block. In other words, bias introduces correlations between justices' voting patterns, which in turn result in increased predictability.

From this perspective, the predictability of a justice with respect to her predictability in an equivalent ideal court provides a quantitative proxy for stable justice correlations, which ultimately reflect a priori attitudes towards the law.

### Stochastic block models for vote prediction

To assess to what extent votes of individual Supreme Court justices are predictable due to stable correlations, we use methods that have been developed to analyze complex social networks. In particular, we adapt a method that is able to uncover unobserved associations between actors (in this case, justices) in complex networks [12]. The method relies on the assumption that justices and cases can be grouped into "blocks," which carry relevant information about justices' voting patterns [13–16]. Unlike previous analyses of coalition formation in the Supreme Court [16], we do not assume a priori which blocks are the most relevant or even that there is a single or a few relevant block structures. Rather, we assume that all blocks of justices and cases are possible in principle, and use a Bayesian approach to correctly average over them.

Consider the voting record $\mathbf{V}^n$ of a court up to case $n$. The voting record can be represented as a matrix whose elements are $V^n_{ij} = 1$ if justice $i$ (with $i=1,2,\ldots,9$) voted in favor of the petitioner in case $j$ (with $j \leq n$), and $V^n_{ij} = 0$ otherwise. We want to predict the vote of a justice in case $n$ (without loss of generality we set this justice to be number 1), given the complete voting record of the court up to case $n-1$ and the votes of the other eight justices in case $n$. We denote this available information (or observation) $\mathbf{V}^{n\setminus 1}$, which is the same as $\mathbf{V}^n$ except that the vote of justice 1 in case $n$, $V^{n\setminus 1}_{1n}$, is not defined.

Let's assume that there is a potentially infinite collection $\mathcal{M}$ of generative models that could plausibly explain the voting record of the court (that is, that could generate observation $\mathbf{V}^{n\setminus 1}$). Then, our expectation for the probability of justice 1 voting in favor of the petitioner in case $n$ is

$$p(V^n_{1n}=1|\mathbf{V}^{n\setminus 1}) = \int_{\mathcal{M}} dM\, p(V^n_{1n}=1|M) p(M|\mathbf{V}^{n\setminus 1}), \quad (1)$$

where $p(V^n_{1n}=1|M)$ is the probability that $V^n_{1n}=1$ in a voting history generated with model $M \in \mathcal{M}$, and $p(M|\mathbf{V}^{n\setminus 1})$ is the plausibility of model $M$ given our observation. Using Bayes theorem, we can rewrite Eq. (1) as

$$p(V^n_{1n}=1|\mathbf{V}^{n\setminus 1}) = \frac{\int_{\mathcal{M}} dM\, p(V^n_{1n}=1|M) p(\mathbf{V}^{n\setminus 1}|M) p(M)}{\int_{\mathcal{M}} dM'\, p(\mathbf{V}^{n\setminus 1}|M') p(M')}, \quad (2)$$

where $p(\mathbf{V}^{n\setminus 1}|M)$ is the probability that model $M$ gives rise to $\mathbf{V}^{n\setminus 1}$ among all possible voting histories, and $p(M)$ is the a priori probability that model $M$ is the one that actually gave rise to $\mathbf{V}^{n\setminus 1}$.

The key to good predictions is to identify sets of models that are general, empirically grounded, and analytically or computationally tractable. We focus on the family $\mathcal{M}_{\mathbf{BM}}$ of stochastic block models [12,19,20]. In a stochastic block model, justices and cases are partitioned into blocks and the probability that a justice votes in favor of the petitioner in a case depends only on the blocks to which the justice and the case belong (see Methods and Ref. [12]).

Under these conditions, the probability that justice 1 votes in favor of the petitioner in case $n$ is:

$$p_{BM}(V_{1n}^n = 1|\mathbf{V}^{n\setminus 1}) = \frac{1}{Z} \sum_{\substack{P_J \in \mathcal{P}_J \\ P_C \in \mathcal{P}_C}} \left(\frac{l_{\sigma_1 \sigma_n} + 1}{r_{\sigma_1 \sigma_n} + 2}\right) \exp[-\mathcal{H}(P_J, P_C)], \quad (3)$$

where the sum is over all possible partitions of the justices and cases into blocks ($\mathcal{P}_J$ and $\mathcal{P}_C$, respectively), $l_{\alpha\beta}$ is the number of votes favorable to the petitioner from justices in block $\alpha$ to cases in block $\beta$, $r_{\alpha\beta}$ is the maximum number of such votes (that is, the number of pairs justice–case such that the justice is in $\alpha$ and the case is in $\beta$), $\sigma_1$ is the block of justice 1 (in partition $P_J$), and $\sigma_n$ is the block of case $n$ (in partition $P_C$).

The weighting function $\mathcal{H}(P_J, P_C)$ depends on the partitions only

$$\mathcal{H}(P_J, P_C) = \sum_{\alpha,\beta} \left[\ln(r_{\alpha\beta} + 1) + \ln\binom{r_{\alpha\beta}}{l_{\alpha\beta}}\right]. \quad (4)$$

Because of the formal analogy of Eq. (3) with a thermal average in statistical physics, one can estimate $p_{BM}(V_{1n}^n = 1|\mathbf{V}^{n\setminus 1})$ using Metropolis sampling [12,21].

## Results

We study the first 150 cases of each of the courts from the first Warren Court (1953) to the last Rehnquist Court (1994–2004). We use the data in the original Supreme Court Database compiled by Spaeth and coworkers [22]. We restrict our analysis to "simple cases" defined as those that verify: (i) all nine justices either voted with the majority or dissented (cases with regular or special concurrence, nonparticipation, or other voting behaviors are excluded); and (ii) were formally decided with full opinions, that is, were granted oral argument and resulted in a signed opinion. For cases dealing with multiple issues, we consider only the main issue, as defined in the database. We restrict ourselves to the first 150 cases of each court because few courts have seen more than 150 simple cases, so results become very noisy after that.

### Justice predictability is higher in real courts than in ideal courts

As discussed above, justice votes in ideal courts are uncorrelated, so the fact that two justices agree or disagree on one case carries no information about their potential agreement on another case. In such a scenario, the stochastic block model cannot possibly extract any useful information from past votes, and we can expect it to be *at most* as accurate as the majority rule. Conversely, in real courts the majority rule is as accurate as in ideal courts because, for this heuristic, the only relevant piece of information is how many justices voted with the majority (for example, in a 6–3 vote, the majority rule will always correctly predict six votes, and incorrectly predict the other three); since the stochastic block model does exploit the block–structure of the voting record, we expect it to be *at least* as accurate as the majority rule in real courts.

Therefore, a "predictability gap" between the majority rule and the stochastic block model in ideal courts would reflect the inability of the latter to capture the little information available in ideal courts; a predictability gap in real courts would reflect the existence of a stable block–structure in the voting record.

To investigate these situations, we artificially generate ideal courts. Given a real court, we generate its corresponding ideal court by randomly reshuffling, within each case, the votes of the justices (Fig. 1). By doing that, each case ends up with the same number of favorable votes as the real case (and, therefore, the same decision), but correlations between justices across cases are eliminated.

For these artificially–generated ideal courts, we find that both the majority rule and the stochastic block model algorithms correctly predict 71% of individual justices votes (Fig. 2A; see Methods for the precise definition of predictability). The absence of predictability gap in ideal courts indicates that, lacking any consistent voting patterns, the stochastic block model is able to capture the maximum possible amount of information.

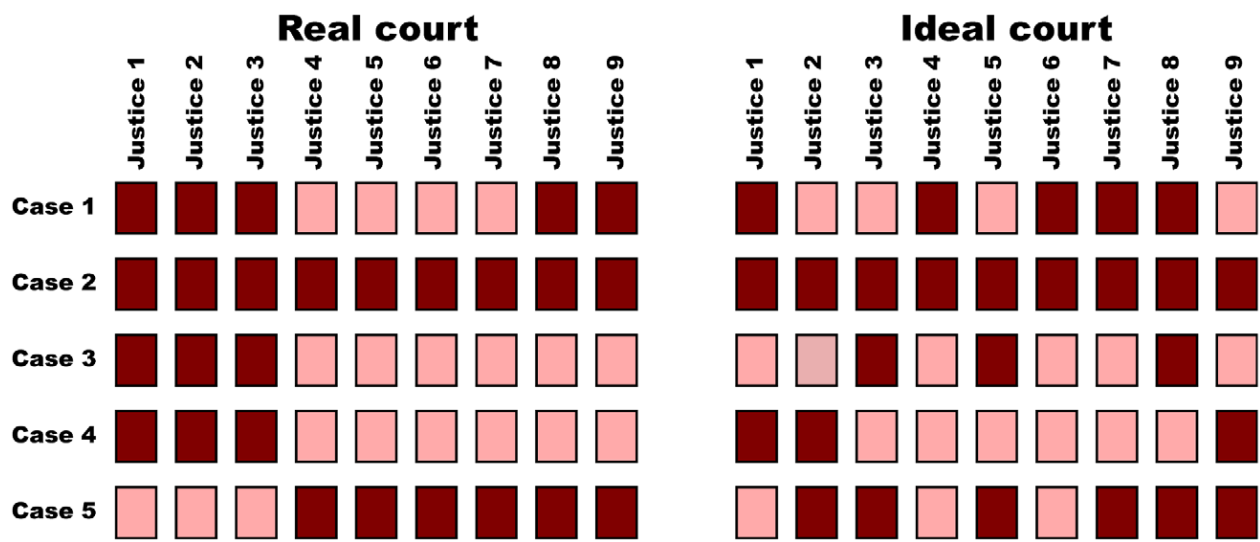

**Figure 1. Court idealization.** Each row represents the votes of the nine justices in a case (dark, agreement with the petitioner; bright, disagreement with the petitioner). We obtain the ideal court (right) from the real court (left) by randomly reshuffling, within each case, the votes of the justices so that the number of agreements and disagreements is preserved.
doi:10.1371/journal.pone.0027188.g001

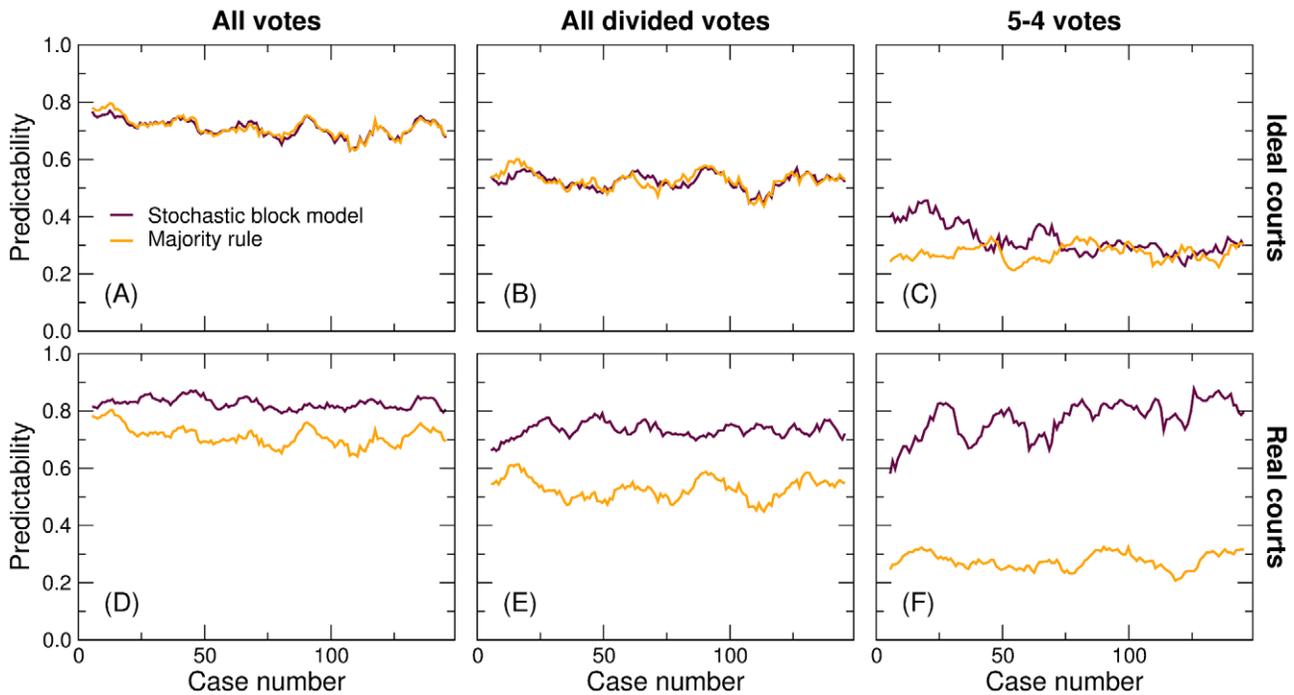

**Figure 2. Average predictability of U.S. Supreme Court votes as a function of the case number (first decision of the court, second decision of the court, and so on; see Methods for the definition of predictability).** Lines represent the moving average (with a window of $\pm 5$ cases) of the average over courts for: (A–C) Ideal courts; (D–F) Real courts. (A, D) Predictability considering all decisions; (B, E) excluding unanimous and 8–1 decisions; (C, F) considering only 5–4 decisions.
doi:10.1371/journal.pone.0027188.g002

For real courts, on the other hand, the block model algorithm correctly predicts 83% of the individual justices' decisions (Fig. 2D), and is therefore consistently more accurate than the majority rule. Although direct comparison with previous forecasting attempts is not straightforward, it is remarkable that our algorithm based on hidden associations between justices is more accurate at predicting justice votes than forecasts made by legal experts as well as those made by algorithms that take into consideration the content of the cases [2,5]. Indeed, while we correctly predict 83% of the votes, legal experts correctly predicted 67.9% and content–based algorithms correctly predicted 66.7% [2,5].

Additionally, one needs to consider that the real extent of the correlations between justices' voting patterns is obscured by two facts. First, for unanimous votes, which constitute a significant fraction of all Supreme Court decisions, both the majority rule and the stochastic block model correctly predict all of the votes. Second, for 8–1 votes both algorithms predict eight votes correctly and one incorrectly. In either case there is no predictability gap.

Since we are interested in non–trivial voting correlations and to eliminate the effect of these somewhat pathologic situations, we turn to what we call ''divided votes,'' that is, votes in which the minority comprises at least two justices (Figs. 2C–D). In ideal courts, divided votes are necessarily more difficult to predict than regular votes: both algorithms accurately predict 53% of the decisions. In real courts, however, the predictability gap widens and the stochastic block model algorithm correctly predicts 73% of the votes.

The widening of the predictability gap becomes even more apparent if we limit our analysis to the cases that are, in principle, most difficult, namely those resulting in a 5–4 vote (Figs. 2E–F). Remarkably, while the majority rule only predicts 28% of these votes correctly, the stochastic block model makes the right prediction in 77% of the cases. This result may appear as a trivial consequence of a single ideological left–right divide in the Supreme Court–that is not the case. Indeed, the most common 5–justice coalition accounts for less than 50% of the 5–4 decisions of the court [16]; our predictions are more accurate thanks to the Bayesian approach that we use to average over all possible justice coalitions and case types.

Besides providing further quantitative evidence for the observation that justices' votes are not immune to the their personal preferences [23–25] and are not independent [17], our approach enables us to quantitatively investigate historical questions about the behaviors of individual justices and courts.

### Relative predictability

The predictability gap in real courts highlights the existence of consistent and stable voting correlations between justices, which our algorithm identifies as justice and case blocks. As we have shown, these correlations become more apparent as one considers closer votes. Based on these observations, we define the relative predictability of a set of votes as the ratio between the predictability of the votes using the stochastic block model, and the predictability *of hypothetical ideal votes in equivalent ideal courts* using the stochastic block model (Methods). A relative predictability of one indicates no deviations from ideal behavior, whereas larger values indicate stable associations between justices.

As discussed above, unanimous decisions and 8–1 votes carry no information about stable associations between justices and, consequently, result in relative predictabilities of 1. Therefore, in the rest of our analysis we consider only divided votes (although all the results reported hold if one considers all the votes instead).

### Justice predictability has decreased over time

Virtually any new nomination to the Supreme Court comes with accusations of judicial bias and/or judicial activism from the opposition party. This raises, among others, the question of whether different justices have different predictabilities and, if so, whether Democrat–nominated justices are more or less predictable than Republican–nominated justices.

We find evidence that not all justices are equally predictable (Fig. 3A). Golberg, the least predictable justice in the period considered is only 2% more predictable than he would have been had he voted ideally in an ideal court (although one needs to consider his short tenure at the Supreme Court). Marshall, on the other extreme, is 64% more predictable than he would have been had he voted ideally in ideal courts. Most justices have relative predictabilities between 1.2 and 1.5 (Fig. 3B). Of note, our measure of predictability may not correlate with expectations based on the liberal or conservative attitudes of the justices. For example, justice Stevens, who is generally regarded as a liberal and was the easiest justice to predict by experts in Refs. [2,5], is relatively unpredictable to the stochastic block model. This means that, liberal or not, for most of his career he did not vote consistently with the same other justices.

We do not find any evidence that Democrat–nominated or Republican–nominated justices are more predictable; in fact, both the most and the least predictable justices were nominated by Democratic presidents (Fig. 3B). Interestingly, however, we do find that justice predictability significantly decreased during the period studied (p = 0.026, Fig. 3C). This global trend seems to be shared by Republican–nominated and Democrat–nominated justices, but is more significant for the former (p = 0.026) than for the latter (p = 0.063).

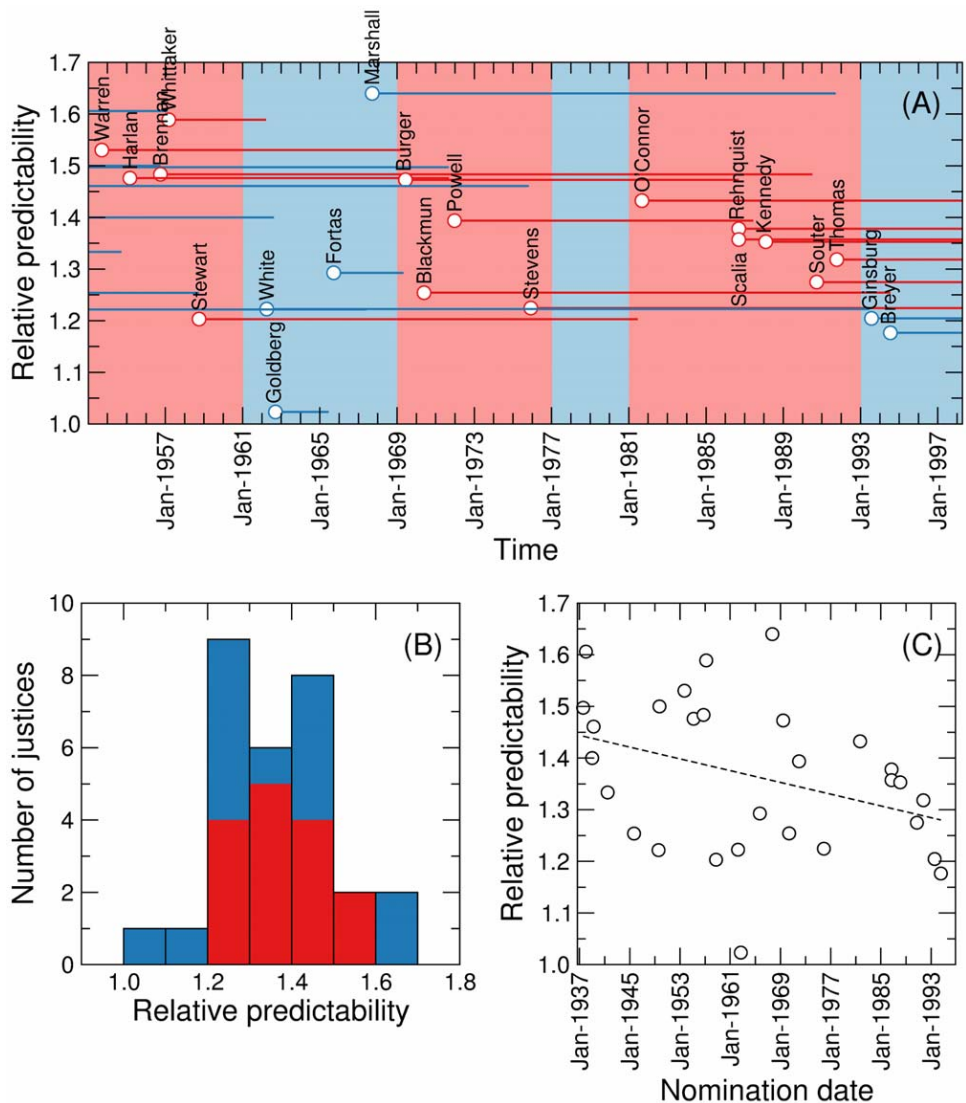

Figure 3. Relative predictability of individual U.S. Supreme Court justices. (A) Each line indicates the average relative predictability (that is, predictability according to the block model algorithm in the real court over the predictability in an equivalent ideal court) of a justice over their tenure, which is indicated by the length of the line. Red lines correspond to Republican–nominated justices and blue lines to Democrat–nominated justices. The background color indicates the party of the president. (B) Histogram of relative predictabilities. Bar colors indicate the fraction of Republican–nominated (red) and Democrat–nominated (blue) justices within each bin. (C) Relative predictability as a function of the nomination date of the judge. Relative predictability has significantly decreased during the period considered (p = 0.026, Spearman's rank correlation), as indicated by the dashed line (which is only shown as a guide to the eye).
doi:10.1371/journal.pone.0027188.g003

## Court predictability is lower during Democratic presidencies

Finally, we investigate the evolution of the relative court predictability, defined as the average relative predictability of the votes of each individual justice for each of the first 150 cases handled by the court (Methods).

As for justices, we find that courts have relative predictabilities significantly different from each other (Fig. 4A). The most predictable courts are over 60% more predictable than their ideal equivalents, whereas the least predictable court (the last Warren court) is only 7% more predictable than its ideal counterpart. Also similar to what happens with individual justices, we find no statistical evidence that courts with more Democrat–nominated justices are more or less predictable than those with more Republican–nominated justices.

In terms of its historical evolution, we find that courts have been significantly less predictable during Democratic presidencies than during Republican presidencies (p = 0.002; Fig. 4B). In fact, the largest predictability drop occurs between the 6th and 8th Warren Courts, coinciding with President Kennedy's election and subsequent assassination.

## Discussion

Predicting justice behavior is a way to test hypotheses about how justices make decisions, so that studying such predictions sheds light in important problems in decision theory and in legal studies and political science. Here we have studied to what extent can one predict the vote of a justice from the votes of other justices in the same case (and the track record of the court). Our approach thus focuses in the stable correlations between justice behaviors, which, we argue, reflect consistent attitudes towards the law.

Our approach complements previous attempts to predict justice behavior from the characteristics of cases alone. In this regard, our predictions turn out to be more accurate than forecasts made by legal experts and by algorithms that take into consideration the content of the cases. We surmise that two main factors explain the success of the approach. First, contrary to heuristic approaches, the Bayesian formalism that we use to account for the information encoded in previous Court decisions is the correct probabilistic treatment of the data for inference of future votes. Second, the block model is more realistic than one-dimensional (liberal-conservative) models often assumed in analysis of the judiciary [16,23,24], which are inspired by classic studies of voting patterns in Congress [26] (where the one-dimensional approximation is likely to be more appropriate and is easier to justify). For example, the block model can account for a justice voting with certain justices on issues of federalism, and with others in issues related to civil rights.

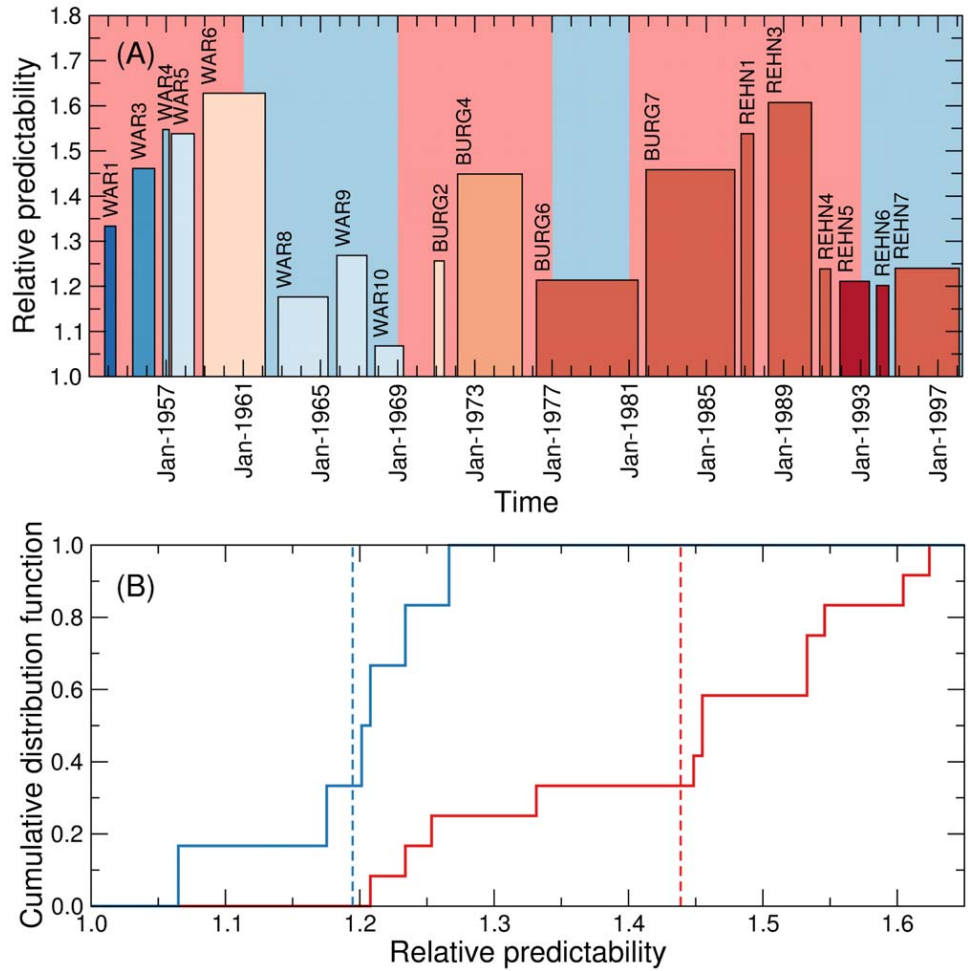

**Figure 4. Relative court predictability.** (A) The height of each bar indicates the average predictability of a court and its width the time span of the court. The color of the bar indicates the makeup of the court, with dark blue corresponding to a court with many Democrat–nominated justices, and dark red to a court with many Republican–nominated justices. (B) Cumulative distribution functions of the relative court predictability for courts that operated mostly under Democratic presidencies (blue) and Republican presidencies (red). The dashed lines indicate the means of the respective distributions.
doi:10.1371/journal.pone.0027188.g004

We have found that justices are significantly more predictable than one would expect from an ideal situation in which justice decisions are uncorrelated. These deviations become more evident for more evenly divided cases; even in 5–4 decisions, which might be expected to be the hardest to predict, our algorithm is able to correctly predict 77% of the individual justices' votes.

Perhaps more importantly, our approach enables us to quantitatively investigate a number of questions related to the historical evolution of the Supreme Court. In particular, we find that justices that were appointed towards the end of the period considered were, in general, less predictable than those that were appointed at the beginning. In addition, we observe that courts operating during Democratic presidencies have been consistently and significantly less predictable than those operating during Republican presidencies.

Providing explanations and contrasting theories for these empirical observations is beyond the scope of this work. In this regard, it is important to note that justice attitudes are only revealed by their votes on cases, so although we focus on justice correlations it is impossible to separate those from correlations between cases. That means, in particular, that predictability may reflect factors that are exogenous to the court, such as the makeup of the cases on which the court has to rule. For example, a court may be less predictable if it has to rule on less politically-charged cases (so justice attitudes play a smaller role), or if it has to rule on very disparate cases (where there is little information from case to case).

In any case, we believe that our approach and our findings open the door for testing such theories in a quantitative manner. More broadly, we believe that out empirical findings illustrate the sort of patterns that quantitative analysis of historical data can help to uncover, and also illustrate how quantitative analysis can help formulate new questions and hypothesis for historical inquiry [27,28].

## Methods

### Outline of the reliability calculations

Formally, a block model $M = (P_J, P_C, \mathbf{Q})$ is completely determined by the partitions $P_J$ and $P_C$ of justices and cases into blocks, and a matrix $\mathbf{Q}$ whose elements $Q_{\alpha\beta}$ represent the probability that a justice in block $\alpha$ votes in favor of the petitioner in a case in block $\beta$. Therefore, Eq. (2) can be rewritten as

$$p_{\text{BM}}(V_{1n}^n = 1 | \mathbf{V}^{n\backslash 1}) = \frac{1}{Z} \sum_{\substack{P_J \in \mathcal{P}_J \\ P_C \in \mathcal{P}_C}} \quad (5)$$

$$\int_{[0,1]^G} d\mathbf{Q} \, p(V_{1n}^n = 1 | P_J, P_C, \mathbf{Q}) p_{\text{BM}}(\mathbf{V}^{n\backslash 1} | P_J, P_C, \mathbf{Q}) \, p(P_J, P_C, \mathbf{Q}),$$

where $\mathcal{P}_J$ (respectively $\mathcal{P}_C$) is the space of all possible partitions of the justices (cases) into blocks, $G$ is the number of distinct block pairs, and $Z$ is a normalizing constant.

Within the family of stochastic block models, one can evaluate the likelihood of each model $M$ because the probability of justice $i$ voting in favor of the petitioner in case $j$ depends only on the blocks to which they belong. We have that [20]

$$p_{\text{BM}}(\mathbf{V}^{n\backslash 1} | P_J, P_C, \mathbf{Q}) = \prod_{\alpha \leq \beta} Q_{\alpha\beta}^{l_{\alpha\beta}} (1 - Q_{\alpha\beta})^{r_{\alpha\beta} - l_{\alpha\beta}}, \quad (6)$$

where $l_{\alpha\beta}$ is the number of votes favorable to the petitioner in $\mathbf{V}^{n\backslash 1}$ between justices in block $\alpha$ and cases in block $\beta$, and $r_{\alpha\beta}$ is the maximum number of such votes (that is, the number of pairs justice–case such that the justice is in $\alpha$ and the case is in $\beta$). Note that we exclude element $(1,n)$ when computing $l_{\alpha\beta}$ and $r_{\alpha\beta}$.

Using that $p(V_{ij}^{n\backslash 1} = 1 | P_J, P_C, \mathbf{Q}) = Q_{\sigma_i \sigma_j}$ (where $\sigma_i$ is the block of justice $i$ in partition $P_J$ and $\sigma_j$ is the block of case $j$ in partition $P_C$) and assuming no prior knowledge about the models (that is, $p(P_J, P_C, \mathbf{Q}) = \text{const.}$), one can use Eqs. (5) and (6) to obtain Eq. (3).

### Predictability definitions

Let $V_{ij}$ be the vote of justice $i$ in case $j$ ($V_{ij} = 1$ if justice $i$ agreed with the petitioner in case $j$, and 0 otherwise), and $V_{ij}^A$ the prediction of algorithm $A$ for that vote. The predictability of a set of $S$ of decisions is

$$p^S = 1 - \frac{1}{\|S\|} \sum_{(i,j) \in S} |V_{ij}^A - V_{ij}| \quad (7)$$

where $\|S\|$ is the number of decisions in the set.

In particular, the predictability $p_i^A$ of justice $i$ is defined as the fraction (over the whole career of the justice) of correctly predicted votes for that justice:

$$p_i^A = 1 - \frac{1}{\|C_i\|} \sum_{j \in C_i} |V_{ij}^A - V_{ij}| \quad (8)$$

where $C_i$ is the set of cases in which $i$ participated.

Similarly, the predictability $p_c^A$ of a court is

$$p_c^A = 1 - \frac{1}{9 \|C_c\|} \sum_{i=1}^{9} \sum_{j \in C_c} |V_{ij}^A - V_{ij}| \quad (9)$$

where $C_c$ is the set of cases heard by court $c$.

We define the relative predictability $r^A$ of an algorithm $A$ as the ratio between the predictability of the real rulings over the same predictability in an equivalent ideal court (obtained as described in Fig. 1):

$$r^A = \frac{(p_c^A)^{\text{real}}}{(p_c^A)^{\text{ideal}}}. \quad (10)$$

## Acknowledgments

We thank L.A.N. Amaral, A. Fernández, and J.M. Mateo for helpful comments and suggestions.

## Author Contributions

Conceived and designed the experiments: RG MS-P. Performed the experiments: RG MS-P. Analyzed the data: RG MS-P. Contributed reagents/materials/analysis tools: RG MS-P. Wrote the paper: RG MS-P.